\documentclass[twocolumn,showpacs,floatfix,preprintnumbers,amsmath,amssymb]{revtex4}
\usepackage{graphicx}
\usepackage{epsfig}
\usepackage{dcolumn}% Align table columns on decimal point
\usepackage{bm}% bold math

\begin{document}

\title{Role of isospin in the nuclear liquid-gas phase transition}
%\title{Influence of isospin on nuclear liquid-gas phase transition}
%\title{Influence of isospin on the order of nuclear liquid-gas phase transition}

\author{C. Ducoin$^{(1,2)}$, Ph. Chomaz$^{(1)}$ and F. Gulminelli$^{(2)}$}

\affiliation{
(1) GANIL (DSM-CEA/IN2P3-CNRS), B.P.5027, F-14076 Caen c\'{e}dex 5, France \\
(2) LPC (IN2P3-CNRS/Ensicaen et Universit\'{e}), F-14050 Caen c\'{e}dex, 
 France
 }

\begin{abstract}
We study the thermodynamics of asymmetric nuclear matter 
using a mean-field approximation with a Skyrme effective interaction, 
in order to establish its phase diagram and more particularly the 
influence of isospin on the order of the transition. 
A new statistical method is introduced to study the thermodynamics
of a multifluid system, keeping only one density fixed, the others 
being replaced by their intensive conjugated variables. In this ensemble,
phase coexistence reduces to a simple one-dimensional Maxwell construction. 
For a fixed temperature under a critical value, a coexistence line is
obtained in the plane of neutron and proton chemical potentials. 
Along this line, the grand potential presents a discontinuous slope, 
showing that the transition is first order except at the two ending 
points where it becomes second order. 
This result is not in contradiction with the already reported 
occurrence of a continuous transformation when a constant proton fraction is 
imposed. Indeed, the proton fraction being an order parameter 
in asymmetric matter, the constraint can only be fulfilled by gradual 
phase mixing along the first-order phase-transition line, leading to 
a continuous pressure.
\end{abstract}

\maketitle
\section{Introduction}
Nucleons in atomic nuclei interact through a finite-range attractive and a
short-range repulsive force. For systems of particles interacting this way,
one expects to find a phase transition analogous to the liquid-gas
transition of a van der Waals fluid \cite{Huang}. 
As a matter of fact, it is recognized that symmetric nuclear matter should
undergo a first-order transition between a low (gas) and a high
(liquid) density phase up to a critical temperature \cite{Finn, Bertsch,Das-PhysRep}. For such
a system containing an equal number of neutrons and protons, isospin
symmetry imposes that the nucleons behave as one single fluid and one
expects a discontinuous density versus pressure equation of state.

The case of asymmetric matter is more complex to study, since there is an
additional degree of freedom to consider: the isospin. Such matter plays an
important role in astrophysics, where neutron-rich systems are involved in
neutron stars and type-II supernova evolution \cite{Lattimer-PhysRep, Glendenning}. 
For asymmetric systems containing a fixed proton fraction, it has been 
shown that the thermodynamic transformations result in a continuous 
evolution of the observables. 
In particular, the system density is a continuous function of the pressure. 
This has been interpreted as the occurrence of a continuous transition \cite{Muller-Serot, Das-PRC, Qian, Sil, Srivastava, Carmona}. 
We will show in the present article that this conclusion is not correct: 
it results from a confusion between the notions 
of 'continuous transition' and 'continuous transformation'. 
Indeed, the phenomenon of isospin distillation demonstrates that the proton 
fraction is an order parameter in asymmetric nuclear matter. Thus, when the 
proton fraction is kept constant, the system is forced to follow the first-order phase-transition line, hiding the discontinuity of the 
thermodynamic-potential first derivative.

The plan of the paper is as follows: 
we calculate in Section II the nuclear matter 
grand potential in the mean-field approximation using a Skyrme (Sly230a) energy 
density functional \cite{Chabanat}. The thermodynamics of nuclear matter has 
been addressed earlier with different effective forces 
\cite{Barranco, Lattimer-PRC, Shlomo, Douchin}. 
The construction of phase coexistence is presented in section III. 
It consists in correcting the mean-field instabilities with the introduction of phase separation, in order to build the concave envelope of the thermodynamic potential. To perform this Gibbs construction with an arbitrary number of conserved quantities, we introduce a new method that reduces this multidimensional problem to a simple one-dimensional Maxwell construction on a carefully defined statistical potential. 
The results of this analysis are presented in section IV.
The grand-canonical potential presents a discontinuous derivative on both sides of a bi-dimensional manifold limited by a critical line in the 3-dimensional space including the temperature and the proton and neutron chemical potentials. The transition is thus first order for all associated proton fractions. Only the critical line corresponds to continuous transitions. An interesting point is that in this 3-dimensional problem the critical line can be characterized by additional critical exponents as the chemical potential approaches its critical value for a given temperature.
We first present the coexistence region obtained for a fixed temperature, stressing the isospin properties of phase coexistence. Then, the thermodynamic consequences of a transformation at constant proton fraction are analyzed. We show that this transformation forces the system to follow the coexistence line, and this is the generic behavior expected when a conservation law acts on an order parameter. In this case, the first-order phase transition results in a continuous transformation from a diluted to a dense system through a phase coexistence, which should not be confused with a continuous transition. 
Indeed, one should clearly make the difference between a transformation, which is a specific path in the space of thermodynamic variables, and a phase transition which is an anomaly of the thermodynamic potential considered in the total space of thermodynamic variables
\footnote{
Phase transitions are related to the properties of the thermodynamic potential partial derivatives with respect to thermodynamic variables (i.e. extensive observables $x_i$ and their intensive conjugates $y_j$). 
%The thermodynamic potential $G(x_i,y_j)$ has then to be studied in the total space of thermodynamic variables, and not along a spectific path, in order to determine the presence and nature of a phase transition. For this reason, transition properties can not be deduced from the properties of a single transformation, which corresponds to a specific path in the space of thermodynamic variables.  
A transformation is defined by a given path $(x_i=f_i(\lambda),y_j=g_j(\lambda))$, where $\lambda$ is a curvilinear coordinate. 
The thermodynamic potential 
total derivative for this transformation is :
\[
\frac{dG}{d\lambda}=\sum_i \frac{\partial G}{\partial x_i} f_i^{\prime}+ \sum_j \frac{\partial G}{\partial y_j} g_j^{\prime}
\]
The properties of this total derivative depend on the specific path, while the transition is characterized by the partial derivatives alone.
	}
.
The temperature dependence of the phase diagram is finally explored, and the critical behaviour is presented.

\section{Thermodynamics with Skyrme forces}

\subsection{Isospin dependent energy functional}

We study the case of nuclear matter in a mean-field approach, with the 
SLy230a Skyrme effective interaction \cite{Chabanat}. 
This local interaction allows to
introduce an energy density $\mathcal{H}{(\mathbf{r})}$ so that
the total energy for a system of nucleons in a Slater determinant 
$\mid \psi>$ reads :
\begin{equation}
\langle \psi | \hat{H}| \psi \rangle=\int {\mathcal{H}(\mathbf{r})d}\mathbf{r}
\label{EQ:meanH}
\end{equation}

where $\hat H$ is the Hamiltonian of the system.
The energy density $\mathcal{H}$ is a functional of the particle
densities $\rho _{q}$ and kinetic densities $\tau _{q}$ for neutrons ($q=n$)
and protons ($q=p$). 
Denoting $\hat{\rho}_{q}$ the one-body density matrix of the
particles of type $q$, those quantities are expressed as follows: 
$\rho _{q}(r)=\langle r | \hat{\rho}_{q} | r \rangle$ 
and 
$\tau _{q}(r)=\langle r | \frac{1}{\hbar ^{2}} \hat{p} \hat{\rho}_{q}\hat{p} 
| r \rangle$, 
such that $\frac{\hbar ^{2}}{2m}\tau _{q}$ is the kinetic energy density.

For later convenience it is useful 
to introduce the isoscalar and isovector densities :
\begin{equation}
\begin{array}{ll}
\rho =\rho _{n}+\rho _{p}\;, & \;\tau =\tau _{n}+\tau _{p} \label{EQ:rho} \\
\rho _{3}=\rho _{n}-\rho _{p}\;, & \;\tau _{3}=\tau _{n}-\tau _{p}
\end{array}
\end{equation}

In the case of homogeneous, spin-saturated matter with no
coulomb interaction, four terms contribute to the energy density :
\begin{equation}
\mathcal{H}=\mathcal{K}+\mathcal{H}_{0}+\mathcal{H}_{3}+\mathcal{H}_{eff} \label{EQ:mathH}
\end{equation}

In this expression, $\mathcal{K}$ is the kinetic-energy term, 
$\mathcal{H}_{0}$ a density-independent two-body term, $\mathcal{H}_{3}$ 
a density-dependent term, and $\mathcal{H}_{eff}$ a momentum-dependent term:
\begin{eqnarray}
\mathcal{K}&=&\frac{\hbar ^{2}}{2m}\tau\\ 
\mathcal{H}_{0} &=&C_{0}\rho ^{2}+D_{0}\rho _{3}^{2}\\
\mathcal{H}_{3} &=&C_{3}\rho ^{\sigma +2}+D_{3}\rho ^{\sigma }\rho _{3}^{2}\\
\mathcal{H}_{eff} &=&C_{eff}\rho \tau +D_{eff}\rho _{3}\tau _{3}
\end{eqnarray}

The coefficients $C_{i}$ and $D_{i}$, associated respectively 
with the symmetry and asymmetry contributions, are linear combinations of the traditional Skyrme parameters :
\begin{equation}
\begin{array}{ll}
C_{0}&= \ \ 3t_{0}/8 \\ 
D_{0}&=- t_{0}(2x_{0}+1)/8 \\ 
C_{3}&= \ \ t_{3}/16 \\ 
D_{3}&=-t_{3}(2x_{3}+1)/48 \\ 
C_{eff}&= \ \ [3t_{1}+t_{2}(4x_{2}+5)]/16 \\ 
 D_{eff}&= \ \ [t_{2}(2x_{2}+1)-t_{1}(2x_{1}+1)]/16
 \label{EQ:SkyrmeCoef}
\end{array}
\end{equation}

\begin{figure}[tbph]
 \begin{center}
 \includegraphics*[height=0.9\linewidth]{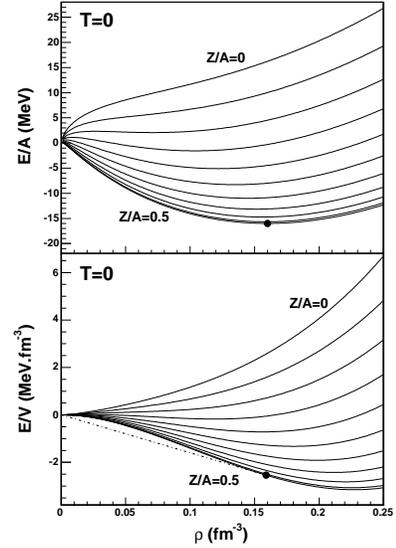}
 \end{center}
 \caption
 	{
Sly230a functional for infinite nuclear matter : 
energy per particle (upper part) and energy density
(lower part) as functions of the total nucleon density for regularly spaced proton
fractions from Z/A=0.5 to Z/A=0, in steps of 0.05.
	}
\label{FIG:FsV-FsA-T0}
\end{figure}

To illustrate the SLy energy functional 
we present in figure \ref{FIG:FsV-FsA-T0} 
the energy density and the energy per particle as a
function of the total particle density for various proton fractions. 
The kinetic-energy term has been computed integrating over the Fermi spheres associated
with the considered proton and neutron densities.
The minimum of the energy per particle is the saturation point of 
symmetric matter $\rho_0=0.16$ fm$^{-3}$ 
and $E_0=-15.99$ MeV,
while the curvature gives the 
incompressibility $K=230.9$ MeV. The $Z/A=0$ curves correspond to 
pure neutron matter which does not saturate, the Sly forces being 
fitted on realistic neutron-matter EOS calculations \cite{Chabanat,Douchin}.

The mean-field effective Hamiltonian 
$\hat{W}_{q}$ for each particle-type q is
defined by the relation 
$\delta \langle \hat{H}\rangle =Tr(\hat{W}_{q}\delta \hat{\rho_{q}})$ 
i.e. as a functional derivative of the energy density. 
In the case of Skyrme interactions, this leads to the expression : 
\begin{equation}
\hat{W}_{q}=\frac{\partial \mathcal{H}}{\partial \tau _{q}}
\frac{\hat{p}^{2}}{\hbar ^{2}}+\frac{\partial \mathcal{H}}{\partial \rho _{q}} 
=\frac{1}{2m_{q}^{*}}\hat{p}^{2}+U_{q} 
\label{EQ:MeanField}
\end{equation}
where $U_{q}=\partial _{\rho _{q}}\mathcal{H}$ is the local mean-field potential :
\begin{eqnarray}
 U_{q} &=&U_{0_{q}}+U_{3_{q}}+U_{eff_{q}} \label{EQ:U} \\
 &=& [2C_{0}\rho 
 			+(\sigma+2)C_{3}\rho ^{\sigma+1}+\sigma D_{3}\rho ^{\sigma-1}\rho_{3}^{2}
			+C_{eff}\tau] \nonumber \\
 &\pm&[2D_{0}\rho_{3}+2D_{3}\rho^{\sigma}\rho_{3}+D_{eff}\tau_{3}] \nonumber
 \end{eqnarray}
and the effective mass $m_{q}^{*}$ is defined by :
\begin{equation}
 \frac{\hbar^{2}}{2m_{q}^{*}} 
 =\frac{\hbar ^{2}}{2m}+C_{eff}\rho \pm D_{eff}\rho _{3} \label{EQ:m*}
 \end{equation}
In both expressions, the $\pm$ sign refers to neutrons $(+)$ or protons $(-)$.

In uniform nuclear matter, the eigenstates of this mean-field 
Hamiltonian are spin-up 
or spin-down plane waves with momenta $\mathbf{p}_{i}$. The single-particle energies
are given by : 
\begin{equation}
\epsilon _{q}^{i}=\frac{p_{i}^{2}}{2m_{q}^{*}}+U_{q} 
\label{EQ:epsilon-i}
\end{equation}

 \subsection{Finite temperature}

Within a mean-field approach, thermodynamic relations are easy to derive in
the grand-canonical ensemble. 
For a system of neutrons and protons, the grand-canonical constraint
imposes the average value of three observables:
energy, number of protons and neutrons. 
Maximizing the Shannon
entropy with these three constraints leads to an equilibrium partition sum :
\begin{equation}
Z_{GC}=Tr[e^{-\beta \hat{H}+\alpha_{n} \hat{N}_{n}+\alpha_{p}\hat{N}_{p})}] \label{EQ:ZGC}
\end{equation}
where the inverse temperature $\beta =1/kT$ is the Lagrange multiplier 
associated with the energy constraint, and $\alpha _{q}=\beta \mu _{q}$ are
the Lagrange multipliers controlling the particle numbers $<\hat N_{q}>$, 
$\mu _{q} $ being the chemical potentials. 
The Lagrange parameters fulfill the equations of state :
\begin{eqnarray}
\langle \hat{H}\rangle &=& -\partial _{\beta }\ln Z_{GC} \\
\langle \hat{N}_{q} \rangle &=& \partial _{\alpha _{q}}\ln Z_{GC}
\end{eqnarray}

The self-consistent mean-field approximation amounts to 
use independent particle states as trial density matrices 
in the maximum-entropy variational principle: 
the single-particle states of
energy $\epsilon _{q}^{i}$ are then occupied 
according to the Fermi-Dirac distribution \cite{Vautherin} : 
\begin{equation}
n_{q}^{i}=\frac{1}{1+exp(\beta (\epsilon _{q}^{i}-\mu _{q}))}
\end{equation}

In infinite matter, $n_{q}^{i}$ is a continuous distribution 
$n_{q}(p)$ and the densities $\rho _{q}$ and $\tau _{q}$ read :
\begin{eqnarray}
\rho _{q} &=&2\int_{0}^{\infty } n_{q}(p) 
\frac{4\pi p^{2}}{h^{3}} dp\label{EQ:density}\\
\tau _{q} &=&2\int_{0}^{\infty } \frac{p^{2}}{\hbar ^{2}} 
n_{q}(p) \frac{4\pi p^{2}}{h^{3}} dp \label{EQ:tau}
\end{eqnarray}
where the factor $2$ come from the spin degeneracy. 

The first equation establishes a self-consistent relation between the
density of q-particles $\rho _{q} $ and 
their chemical potential $\mu _{q}$. The above densities 
can be written as regular Fermi integrals by shifting the chemical potential
according to $\mu _{q}^{\prime }=\mu _{q}-U_{q}.$ 
The Fermi-Dirac distribution indeed reads :
\begin{equation}
n_{q}(p)=\frac{1}{1+exp(\beta (p^{2}/2m_{q}^{*}-\mu _{q}^{\prime }))}
\label{EQ:distribution}
\end{equation}

Equations (\ref{EQ:distribution}) and (\ref{EQ:density}) 
define a self-consistent problem since $m_{q}^{*}$ depends 
on the densities according to eq.(\ref{EQ:m*}). 
For each pair $(\mu _{n}^{\prime },\mu _{p}^{\prime })$
a unique solution $(\rho _{n},\rho _{p})$ is found by iteratively
solving the self-consistency between $\rho _{n,p}$ and $m_{n,p}^{*}$. 
Then eq.(\ref{EQ:tau}) is used to calculate $\tau_{n,p}$. 
These quantities allow to compute the 
one-body partition sum :
\begin{equation}
Z_{0}=Tr[e^{-\beta (\hat{W}_{n}+\hat{W}_{p}-\mu _{n}\hat{N}_{n}-\mu _{p}
\hat{N}_{p})}]=Z_{0}^{n}Z_{0}^{p}
\end{equation}
where each partition sum $Z_{0}^{q}$ can be expressed 
as a function of the corresponding kinetic energy density : 
\begin{equation}
\frac{\ln Z_{0}^{q}}{V}=2 \int_{0}^{\infty }
\ln(1+e^{-\beta (\frac{p^{2}}{2m_{q}^{*}}-\mu _{q}^{\prime })})
\frac{4\pi p^{2}}{h^{3}} dp
=\frac{\hbar^2}{3m_{q}^{*}}\beta \tau _{q}
\end{equation}

At the thermodynamic limit, the system volume $V$ diverges 
together with the particle numbers $\langle \hat{N}_{q} \rangle$, 
and the thermodynamics is completely defined
as a function of the two particle densities $\rho_n,\rho_p$.

We can now use the maximum-entropy variational principle to evaluate the
mean-field approximation to the grand-canonical partition sum $Z_{GC}$. 
We recall that the exact grand-canonical ensemble corresponds to the 
maximum of the constrained Shannon entropy which is nothing but $\ln Z_{GC}$ :
\begin{equation}
\ln Z_{GC}=S_{GC}-\beta 
(\langle \hat{H}\rangle_{GC}-\mu _{n}
 \langle \hat{N}_{n}\rangle_{GC}-\mu _{p}
 \langle \hat{N}_{p}\rangle_{GC}). 
\end{equation} 

The variational principle thus states that the mean-field constrained entropy 
is the best approximation within the ensemble of independent-particle trial
states to the exact maximum $\ln Z_{GC}$:
\begin{equation} 
\ln Z_{GC}\simeq S_{0}-\beta 
(\langle\hat{H}\rangle_{0}-\mu_{n}
 \langle\hat{N}_{n}\rangle_{0}-\mu_{p}
 \langle\hat{N}_{p}\rangle_{0}), 
\end{equation}
where the mean-field energy and particle numbers are
defined as functions of densities, i.e. single-particle occupations
eq.(\ref{EQ:distribution}), by :
\begin{equation}
\langle\hat{H}\rangle_{0}=V\mathcal{H} \; ; \; 
\langle\hat{N}_{q}\rangle_{0}=V \rho_q
\end{equation} 

The mean-field entropy is given by :
\begin{equation} 
S_{0}=\ln Z_{0}+\beta 
(\langle\hat{W}\rangle_{0}-\mu_{n}
 \langle\hat{N}\rangle_{0}-\mu_{p}
 \langle\hat{N}_{p}\rangle_{0}) 
\end{equation}
where $\langle\hat{W}\rangle_{0}$ 
is the average single-particle energy :
\begin{equation}
\langle\hat{W}\rangle_{0}=
2V \sum_q \int_{0}^{\infty }
n_q(p) e_q(p)
\frac{4\pi p^{2}}{h^{3}} dp
=-\partial_\beta \ln Z_0
\end{equation}
with $e_q(p)=p^{2}/2m^*_{q} +U_{q}$.

The grand-canonical partition sum in the mean-field approximation
is thus modified with respect to the independent-particle partition sum $Z_0$
as :
\begin{equation}
\ln Z_{GC}\simeq \ln Z_{0}+\beta \left( 
\langle\hat{W}\rangle_{0}-\langle\hat{H}\rangle_{0}\right) 
\end{equation}
which allows to express the grand-canonical potential 
density as a function of densities : 
\begin{equation}
-g=\frac{\ln Z_{GC}}{\beta V}\simeq 
\frac{2}{3}\mathcal{K} +
\mathcal{H}_{0}+
(\sigma +1)\mathcal{H}_{3} +
\frac{5}{3}\mathcal{H}_{eff} 
\end{equation}

At the thermodynamic limit, this quantity 
is equivalent to the system pressure $P=-g$. 
Because of ensemble equivalence we can then evaluate 
all the thermodynamic potentials. For example, the canonical 
partition sum, or equivalently the free energy per unit volume f, 
is defined through the Legendre transform : 
\begin{equation}
f=-\frac{\ln Z_{C}}{\beta V}
=g+\mu _{n}\rho _{n}+\mu _{p}\rho _{p} 
\end{equation}

\begin{figure}[tbph]
 \begin{center}
 \includegraphics*[height=0.7\linewidth]{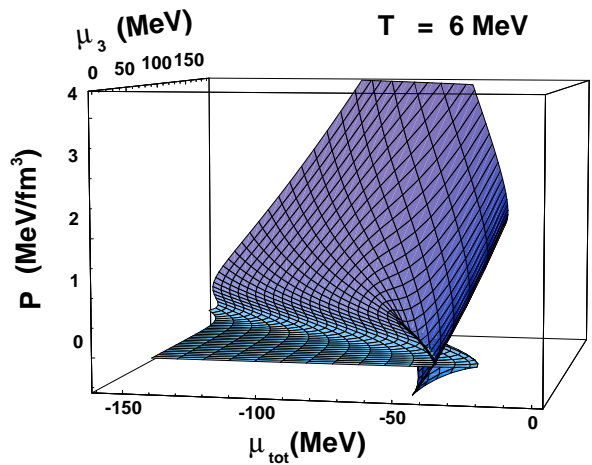}
 \end{center}
 \caption
 	{
Pressure as a function of the isoscalar and isovector chemical
potentials for a uniform system at $T=6 MeV$.
	}
 \label{FIG:MM3P}
 \end{figure}

\begin{figure}[tbph]
 \begin{center}
 \includegraphics*[height=0.7\linewidth]{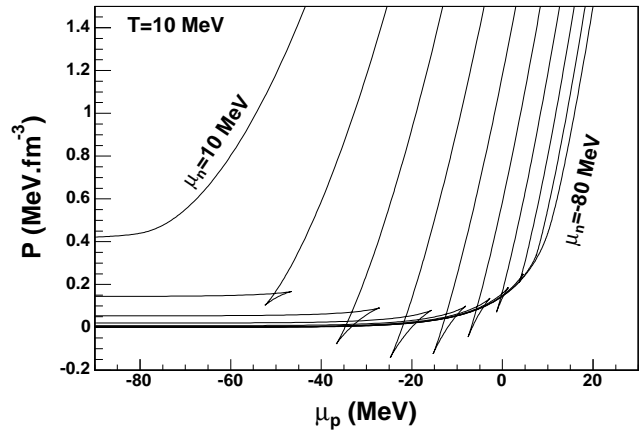}
 \end{center}
 \caption
 	{
Pressure as a function of $\mu_{p}$ for a uniform system at a fixed temperature $T=10 MeV$ and regularly spaced values of $\mu_{n}$ from -80 to 10 MeV.
	}
 \label{FIG:MPP}
 \end{figure}

Figure \ref{FIG:MM3P} illustrates the typical behavior of the pressure 
as a function of the isoscalar and isovector
chemical potentials $\mu =\mu _{n}+\mu _{p}$ and 
$\mu _{3}=\mu _{n}-\mu _{p}$. 
This figure is computed for a uniform system at a fixed temperature. 
For symmetry reasons only the positive $\mu _{3}$ are shown.
We can see that for some values of $(\mu,\mu _{3})$, 
there are three solutions corresponding to different values for 
the conjugated observables $(\rho,\rho _{3})$. 
This is the phase-transition region. 
The true equilibrium is the solution minimizing
the grand potential i.e. maximizing the pressure. 
Thus only the upper part of the pressure manifold 
corresponds to a thermodynamic equilibrium. 
At the resulting fold, the slope changes discontinuously, i.e. using
$\rho_q=\partial g/\partial \mu_q$,
the equilibrium uniform system jumps from a 
low to a high density solution. 
It is a liquid-gas first-order phase transition.
The fold of the grand potential is more clearly shown in figure \ref{FIG:MPP} which presents the pressure as a function of $\mu_{p}$ for different values of $\mu_{n}$.

\section{Phase coexistence}

\subsection{Phase-coexistence conditions in multi-fluid systems}

For systems at the thermodynamic limit, the existence and the order of a phase
transition are intrinsically related to
the singularities of the thermodynamic potential, 
$-T\ln Z\left( \mathbf{\lambda }\right) $ where 
$Z\left( \mathbf{\lambda }\right) $ is the partition sum for
a given macroscopic state characterized by the $L$ intensive parameters 
$\mathbf{\lambda }=\{\lambda _{\ell }\}$ \cite{Balian}. 
These intensive parameters are the Lagrange multipliers introduced in the maximization of the Shannon entropy under the $L$ constraints $<\hat{A}_{\ell }>$ associated with all the $L$
relevant observables $\hat{A}_{\ell }$. 

A system presents a first-order phase 
transition if one of the first partial-derivatives of 
$\ln Z\left( \mathbf{\lambda }\right) $ shows a 
discontinuity \cite{Huang}.
If a non-analyticity (discontinuity or divergence) is present
at a higher order derivative, there is a continuous transition.
Since the equations of states (EOS) are the $L$ relations 
$<\hat{A}_{\ell }>\left( 
\mathbf{\lambda }\right) 
=-\partial _{\lambda _{\ell }}\ln Z\left( \mathbf{\ 
\lambda }\right)$ 
\cite{Balian}, 
we can see that a first-order phase
transition corresponds to a discontinuity in at least one of the EOS, i.e. 
to a jump in the mean value of at least one observable at 
the transition point. 
This observable can then be identified with an order parameter,
since its value allows to distinguish the two coexisting phases.

At the thermodynamic limit, the entropy $S$, as all the extensive variables 
$\{A_{k}\},$ is additive and thus scales like the volume $V$ : 
\begin{equation}
S(V,\{A_{k}\})=V\mathrm{s}(\rho_k=\{A_{k}/V\}) 
\end{equation}

The volume can thus be eliminated from the thermodynamic description
while the other extensive variables can be 
expressed as densities $\mathbf{\rho}=\{\rho_{k}\}$. 
If we consider two isolated systems of volum $V_1=\alpha V$ and 
$V_2=(1-\alpha) V$, 
their constrained entropy is simply the sum of the two
entropies $S=S_{1}+S_{2}$. When $1$ and $2$ are put into contact 
to form a system with
$\mathbf{\rho}=\alpha \mathbf{\rho}_{1} + (1-\alpha) \mathbf{\rho}_{2}$, equilibrium is reached by maximization of the global entropy $S$ so that :
\begin{equation}
S\geq S_{1}+S_{2}\;. 
\end{equation}

This imposes the convexity of the entropy :
\begin{equation}
\mathrm{s}(\alpha \mathbf{\rho}_{1}+(1-\alpha )\mathbf{\rho}_{2})\geq
\alpha \mathrm{s}(\mathbf{\rho}_{1})+(1-\alpha )
\mathrm{s}(\mathbf{\rho}_{2}) 
\end{equation}

As a result, if the homogeneous system has a constrained entropy with a
convex region, a linear interpolation between 
two densities $\mathbf{\rho}_{A}$ and $\mathbf{\rho}_{B}$
corresponds to a physical phase mixing of the two associated states. 
Such linear interpolations lead to the construction of a concave envelope which maximizes the entropy functional
and suppresses the curvature anomaly
(Gibbs construction).
The straight lines corresponding to phase coexistence are 
defined by two points of same tangent plane, 
i.e. with identical first derivatives or intensive variables 
$\lambda _{k}=\partial_{\rho_{k}}\mathrm{s}$, 
and equal values of the constrained entropy, i.e. equal
distances between the entropy $\mathrm{s}(\mathbf{\rho})$ 
and the plane $\sum_k \lambda _{k}\rho_{k}=0$. 
Using $[\partial_{V}S(V,\{A_{k}\})]/V=\mathrm{s}(\{\rho_{k}\})
- \sum_k \lambda _{k}\rho_{k}$ this latter condition can be 
interpreted as the equality of the two system pressures. 
This equality of all intensive variables defines the conditions of
phase equilibrium.

\subsection{Gibbs and Maxwell constructions}

\begin{figure}[tbph]
\begin{center}
\includegraphics*[height=0.7\linewidth]{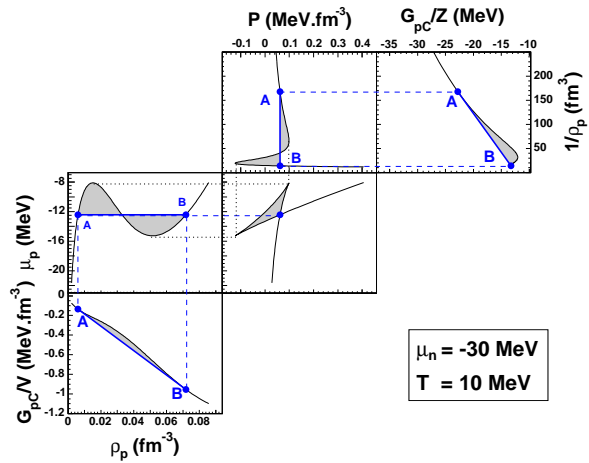}
\end{center}
\caption
	{
Illustration of a Maxwell-Gibbs construction in the 
proton-canonical and neutron-grand-canonical ensemble 
at a fixed neutron chemical potential $\mu_{n}=-30MeV$ and temperature 
$T =10 MeV$. The bottom 
part presents the thermodynamic potential density $G_{pC}/V$ as a function 
of the proton density $\rho_p$. The straight line shows the convex 
envelop interpolating between the two phases $A$ and $B$. 
This corresponds to the Maxwell construction on $\mu_p$ 
as a function of $\rho_p$ (left center). 
The top-right part shows the convex envelop of the thermodynamic potential 
per particle $G_{pC}/Z$ as a function of the inverse of the proton density 
$1/\rho_p$ 
which is associated with a Maxwell construction for the pressure 
as a function of $1/\rho_p$
(top center). 
The two Maxwell constructions correspond to the crossing of the 
$P$-$\mu_p$ diagram (central figure).
	}
\label{const}
\end{figure}

For nuclear matter at a given temperature, 
the $A_{k}$ are the proton and neutron numbers: 
$N=\rho _{n}V$ and $Z=\rho _{p}V.$
Finding two points in equilibrium means finding two sets of densities 
$\{\rho _{n}^{A},\rho _{p}^{A}\}$ and 
$\{\rho _{n}^{B},\rho _{p}^{B}\}$ which fulfill the $3$ equations 
$\mu _{n}^{A}=\mu _{n}^{B}$, $\mu _{p}^{A}=\mu 
_{p}^{B}$, 
$P^{A}=P^{B}$, 
the equality of the temperatures being insured by the use 
of an isothermal ensemble.
The path corresponding to a constant $\mu _{q}$ (e.g. $\mu _{n}$) 
is a curve in the $(\mu _{p},P)$ plane that 
can be determined numerically. 
The problem is now reduced to two dimensions:
if this path shows a crossing point, there are two sets of extensive 
observables for which the intensive parameters are all equal, 
which is the condition for coexistence.
This is illustrated in the central part of figure \ref{const}. 

From the thermostatistics point of view, using the set of state variables
$(\beta,\mu _{n},\rho_p)$ corresponds to 
defining a neutron-grand-canonical but proton-canonical ensemble, 
denoted as $pC$ in the following. The associated potential
per unit volume, $G_{pC}/V=g_{pC}$,
is given by :
\begin{equation}
g_{pC}(\beta ,\mu _{n},\rho _{p})
=\mathcal{H}-\mu_{n}\rho _{n}-s(\mathcal{H},\rho _{n},\rho _{p})/\beta 
\end{equation}
where $s=S/V$ is the entropy density. 
$g_{pC}$ is linked by Legendre transforms both to the
grand-canonical potential $g_{GC}=-P$ (pressure) 
 and to the canonical potential $f$ (free energy) :
\begin{equation}
g_{pC}=-P+\mu_{p}\rho_{p}=f-\mu _{n}\rho_{n}.
\end{equation}

It is important to stress that ensembles are equivalent at the thermodynamic limit which is used in the presented mean-field approach. The difference comes from the thermodynamical variables and
potentials that are used. Therefore, the construction of the phase
transition is not the same. The proposed ensemble leads
to a simple 1-dimensional Maxwell construction which is much simpler
than the usual multidimentional Gibbs construction in a
multi-fluid system. Namely, for the considered system:
\begin{itemize}
\item In the canonical ensemble, the variables are the densities
 ($\rho_{n}$ and $\rho_{p}$) and the potential is the free energy
 f. The Gibbs construction is then a complex 2-dimensional construction
 of its convex envelope.
\item In the grand-canonical ensemble, the variables are the
 chemical potentials ($\mu_{n}$ and $\mu_{p}$) and the potential is
 the grand potential $g$. In
 a first-order phase-transition region this function is multivalued.
 The Gibbs construction is then a complex 2-dimensional construction of the
 intersection of two of the three manifolds.
\item In the proposed ensemble, the variables are one density
 (e.g. $\rho_{p}$) and one chemical potential ($\mu_{n}$) and the potential $g_{pC}$ is the corresponding
 Legendre transform of $f$ and $P$. 
Since this ensemble has only one density left, the phase coexistence is reduced to a simple problem like in a 1-component fluid. At constant $\mu_{n}$, $g_{pC}$ as a function of $\rho_{p}$ should be replaced by its convex envelope. This is now a 1-dimensional problem which is in fact the usual Maxwell construction.
 \end{itemize}

Numerically, we can directly perform this Maxwell construction
on the function $\mu _{p}(\rho _{p})$ for constant $\beta $ and $\mu _{n}$.
The comparison between the results obtained using this method and the method
described in the previous section using the $\mu _{p}-P$ 
crossing point gives an estimation of our numerical error,
which comes out to be less than $0.2\%$ for all temperatures. 

We can additionally remark that the above reasoning also holds 
if we study the $pC$-potential per proton leading 
to a Maxwell construction for the pressure $P$ as a function $1/\rho _{p}$. 
Both constructions are illustrated on figure \ref{const}. 
The convex envelop of $G_{pC}/V$ ($G_{pC}/Z$) corresponds to an
equal-area Maxwell construction on $\mu _{p}(\rho _{p})$ ($P(1/\rho _{p})$) 
and to the crossing point between the two phases in the $\mu _{p}$ 
versus $P$ graph.

It should be noticed that the introduction of a 
statistical ensemble in which only one density is kept, all the other ones 
being replaced by their associated intensive parameters, 
can always be applied to study a phase
transition with a one-dimensional order parameter. 
The only condition is that the considered density 
is not orthogonal to the order parameter, 
i.e. its value is different in the two phases. 
In such an ensemble, the multidimensional Gibbs construction 
reduces to a simple one-dimensional Maxwell construction.

\section{Results}
\subsection{Coexistence region}

 \begin{figure}[tbph]
 \begin{center}
 \includegraphics*[width=0.80\linewidth]{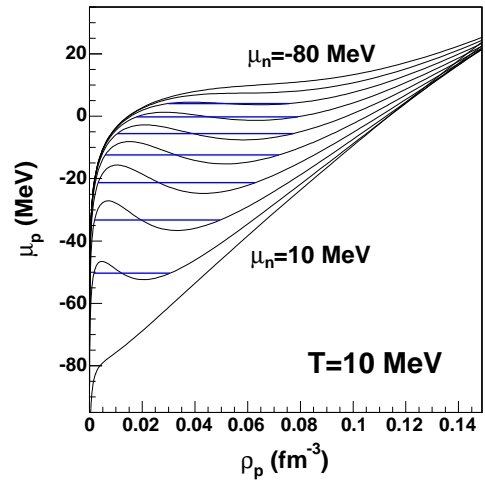}
 \end{center}
 \caption
 	{
 Maxwell construction in the $\mu_p$ versus $\rho_p$ 
 diagram for a fixed temperature $T=10MeV$ and regularly spaced 
 values of $\mu _{n}$ from $-80$ to $10$ MeV.
 	}
 \label{Fig:mu}
 \end{figure}

 In order to explore the phase diagram, the method illustrated in the previous 
 section has to be performed for various temperatures and chemical potentials. 
 Figure \ref{Fig:mu} illustrates the $\mu _{n}$ dependence for $T=10$ MeV. 
 We can see that for a broad range of $\mu _{n}$ the $\mu _{p}(\rho _{p})$ 
 equations of state present a back-bending 
 associated with an instability
 which must be corrected using a Maxwell construction. 
 This defines the transition points $\mu _{p}^{t}(\mu _{n},T)$
 and $P^{t}(\mu _{n},T)$. The ensemble of the transition points 
 constitutes a coexistence curve at the considered temperature. 
 This curve is limited by 
 two critical chemical potentials, $\mu _{n}^{<}(T)$ and $\mu _{n}^{>}(T)$, 
 which in turn define the proton critical 
 chemical potentials 
 $\mu _{p}^{>}(T)=$ $\mu _{p}^{t}(\mu _{n}^{<}(T),T)$
 and 
 $\mu _{p}^{<}(T)=$ $\mu _{p}^{t}(\mu _{n}^{>}(T),T)$,
since $\mu _{p}$ is maximum when $\mu _{n}$ is minimum.
The transition is observed only in a finite range of temperatures 
below a given temperature $T_{c}$ which is nothing but 
the critical temperature of symmetric matter.

\begin{figure}[tbph]
\begin{center}
\includegraphics*[height=0.55\linewidth]{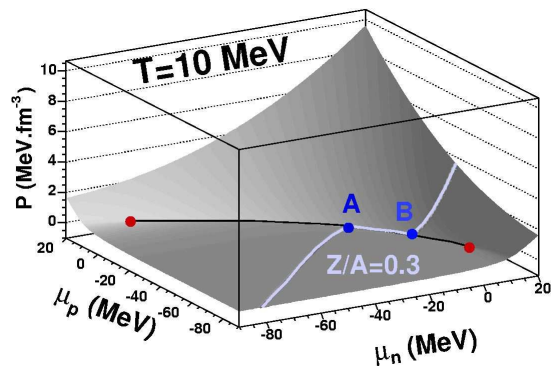}
\end{center}
\caption
	{
Equilibrium pressure computed at a temperature 
 $T=10 MeV$ as a function of $\mu _{n}$ and $\mu_{p}$ 
 after performing a Gibbs construction for different 
 values of $\mu _{n}$. 
 The resulting fold line is the coexistence line ending at
 two critical points. 
The traced path corresponds to a
transformation at constant proton fraction $Z/A=0.3$ (see Section IV-B).
	}
\label{MnMpP}
\end{figure}

The Gibbs construction of phase coexistence leads to well defined
partition sums fulfilling the thermodynamic stability requirement. 
The resulting pressure $P=T\ln Z_{GC}(T,\mu_{n},\mu _{p})/V$ 
at the temperature $T$ $=10$ MeV is shown in figure \ref{MnMpP}.

Along the coexistence line $\mu _{p}^{t}(\mu _{n},T)$, 
the pressure presents a fold, i.e. the derivative perpendicular 
to the line is discontinuous. It is
by definition a region of first-order phase transition. 
The coexistence line at fixed temperature is limited 
by two points 
$(\mu_{n}^{<}(T),\mu _{p}^{>}(T))$ 
and 
$(\mu_{n}^{>}(T),\mu _{p}^{<}(T))$. 
They correspond to vertical tangents in
the grand-potential first derivatives $\rho _{q}(\mu _{n},\mu _{p})$, 
which are singularities in its second derivatives. 
Hence, the limiting points are second-order critical points, 
i.e points of continuous transition.

\begin{figure}[tbph]
\begin{center}
\includegraphics*[height=0.55\linewidth]{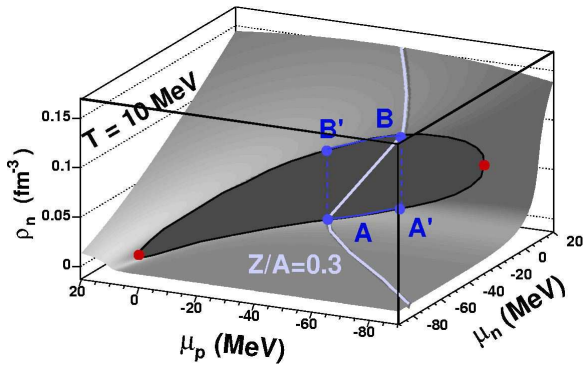}
\end{center}
\caption
	{
First derivative of the grand-canonical partition sum,
 $\rho _{n}=\partial p/\partial \mu _{n}$, as a function of $\mu _{n}$ and 
$\mu_{p}$ at a temperature $T=10 MeV$. The path $Z/A=0.3$ is
also shown.
	}
\label{MnMprn}
\end{figure}

We have represented in figure \ref{MnMprn}
 the first derivative of $\ln Z_{GC}$ in the $\mu _{n}$
 direction, i.e. the neutron density 
 $\rho _{n}=\partial _{\mu _{n}}p$, which is discontinuous 
 on the first-order line and continuous with a 
 vertical tangent at the two critical points. 
Because of the exact isospin symmetry of the SLy interaction, 
the proton density $\rho _{p}$ is symmetric to $\rho _{n}$: 
 it is the same surface with inversion of the axes $\mu _{n}$ and $\mu _{p}$.

\begin{figure}[tbph]
\begin{center}
\includegraphics*[width=0.9\linewidth]{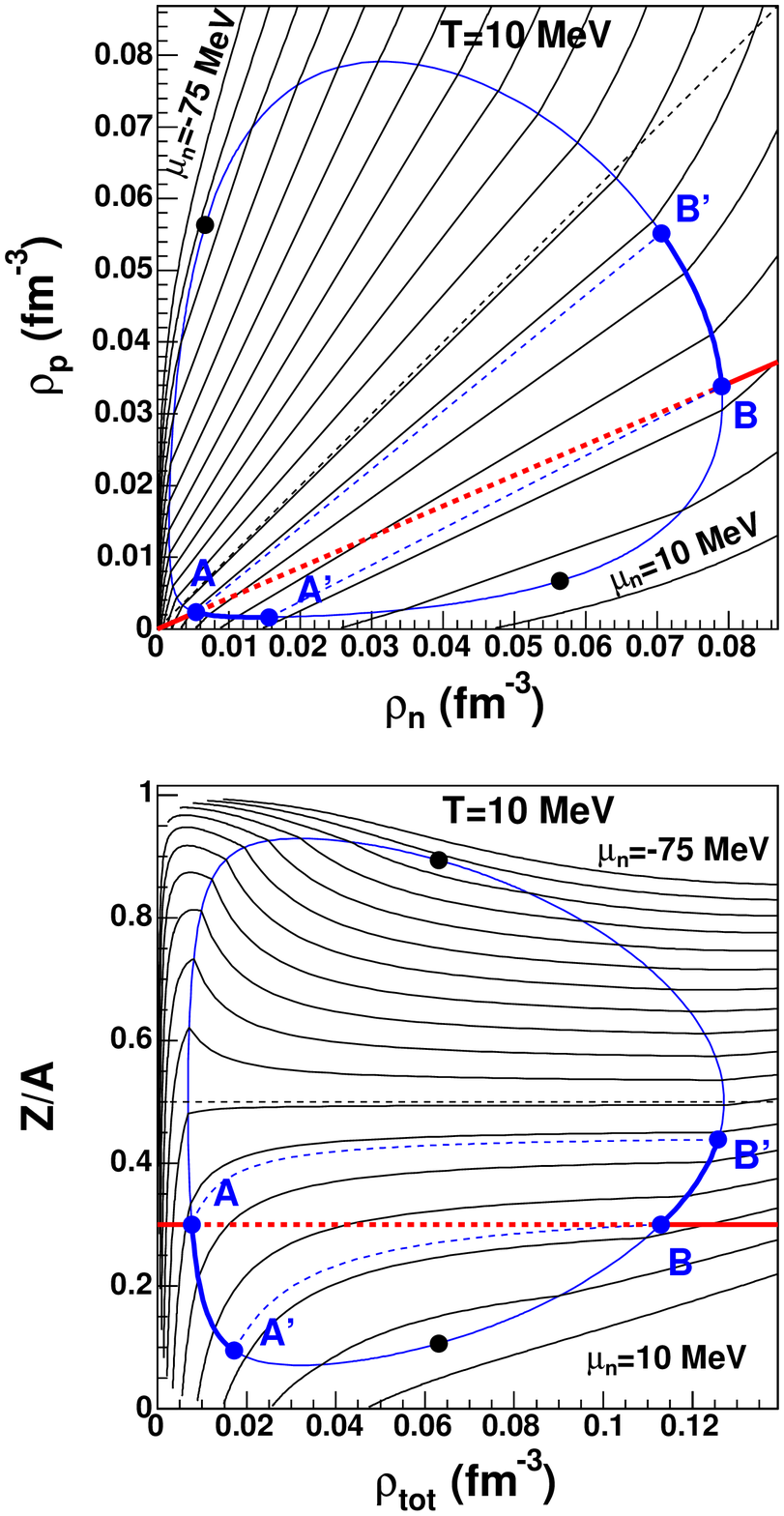}
\end{center}
\caption
	{
Closed curves : coexistence border 
in the proton and neutron-density plane (upper part) and 
total-density and proton-fraction plane (lower part). 
Black dots : critical points.
Full lines : constant $\mu _{n}$ paths 
for regularly spaced 
values between $\mu _{n}=10$ MeV and $-75$ MeV. 
Paths $AA'$ and $BB'$ refer to a transformation at $Z/A=0.3$ (see Section IV-B).
	}
\label{rnrp-rZsA-isoMn}
\end{figure}

The first derivatives of $\ln Z_{GC}$ on both sides of the phase transition
line define pairs of points $(\rho_n,\rho_p)$ 
in coexistence respectively at low (gas) and
high (liquid) density. These phases merge together at the critical points. 
The coexistence region in the proton and neutron-density space is shown on
figure \ref{rnrp-rZsA-isoMn} for a fixed temperature.
The solid lines in this figure give several iso-$\mu_{n}$ paths. 
Because the construction of the concave envelop of the constrained
entropy is nothing but a linear interpolation between two phases, 
inside the coexistence region the iso-$\mu _{n}$ lines are straight lines in 
$(\rho_{n},\rho _{p})$ representation.
The bottom part of the figure shows the coexistence region in the 
total-density and proton-fraction plane $(\rho,Z/A)$, 
which consists in the change of variables 
$\rho =\rho _{n}+\rho _{p}$, 
$y=Z/A=\rho _{p}/(\rho _{n}+\rho _{p})$. 
Because of the non-linearity of the variable change, 
we can see that coexistence does not correspond to a straight line 
in this representation, 
and the value of $Z/A$ evolves when passing from the dense phase
to the diluted phase in coexistence. The only exception is the case of
symmetric nuclear matter, at $y=Z/A=0.5$.

\begin{figure}[tbph]
\begin{center}
\includegraphics*[height=0.55\linewidth]{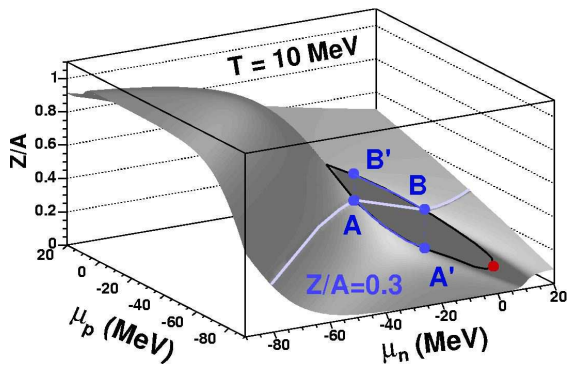} 
\end{center}
\caption
	{
$Z/A=\rho_p/(\rho_n+\rho_p)$ as a function of
$\mu_n$ and $\mu_p$ for a fixed temperature $T=10 MeV$.
 The path at $Z/A=0.3$ is also shown.
 	}
\label{MnMpZsA}
\end{figure}

 The proton-fraction difference in the two phases can be appreciated 
 from figure \ref{MnMpZsA} which shows $Z/A$ as a function of $\mu_n$ 
 and $\mu_p$ for a fixed temperature $T=10 MeV$. 
 $Z/A$ being a combination of the two order parameters $\rho_n$ and $\rho_p$,
 it also presents a discontinuity at the first-order phase-transition border,
 with the only exception of the symmetric nuclear matter point where
 $Z/A=cst=0.5$.
 Correlating this plot with figure \ref{MnMprn},
 one can see that the dense phase (e.g. point B' or B) 
 is systematically closer to isospin symmetry $Z/A=0.5$ 
 than the diluted one (e.g. point A or A'). 
 This phenomenon is known as isospin distillation \cite{Muller-Serot, Baran-NPA, BaoAnLi}. 

\subsection{Transformation at constant $Z/A$}

In low-energy heavy-ion collisions, the proton and neutron numbers
obey two independent conservation laws, implying that 
the proton fraction $Z/A$ is conserved in the reaction.
It is therefore of interest to consider a transformation 
at constant $Z/A$ \cite{Muller-Serot,BaoAnLi}. 
In the ($\rho_n$,$\rho_p$) plane (or equivalently in the 
($\rho_{tot}$,$Z/A$) representation) 
 $Z/A=cst$ transformations are straight lines 
which cross the different constant-$\mu_n$ curves.
Inside the coexistence region, the system with a given proton fraction 
is decomposed into two phases, located at the intersections 
of the coexistence curve
with the corresponding constant-$\mu_n$ curve. 
Since the constant-$\mu_n$ curves are not aligned on constant-$Z/A$ 
lines (except for symmetric matter), 
the constant-$Z/A$ transformation 
does not make a transition from liquid 
to gas at a unique value of $\mu_n$ 
but shows a continuous smooth evolution 
of the intensive parameters along the coexistence line. 

\begin{figure}[tbph]
 \begin{center}
\includegraphics*[height=0.6\linewidth]{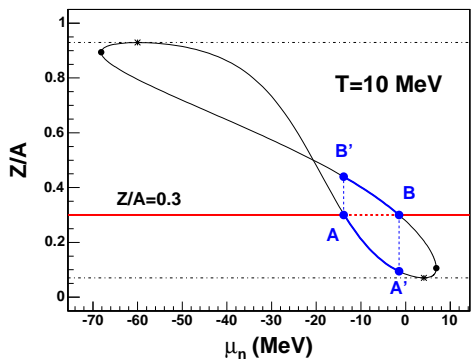} 
\end{center}
\caption
	{
Projection of the coexistence region in the $Z/A$ versus $\mu_n$ plane. 
Black dots are critical points, and stars are points of maximal asymmetry.
The path at $Z/A=0.3$ is also shown.
	 }
 \label{MnZsA}
 \end{figure}

The transformation $Z/A=0.3$ is given as an example by the 
dotted lines in figure \ref{rnrp-rZsA-isoMn}, and the grey path
in figures \ref{MnMpP},\ref{MnMprn},\ref{MnMpZsA}. 
Let us follow this transformation from the low density
phase. 
When the system reaches the coexistence border (point $A$),
a liquid phase appears in $B'$ at the same value of $\mu_n$, $\mu_p$ and $T$.
This point coincides with $A$ in the representation of figure \ref{MnMpP}.
We can see from figures \ref{rnrp-rZsA-isoMn} and \ref{MnMpZsA}
that the liquid fraction is closer to symmetric nuclear matter than 
the original system, as expected from the isospin-distillation phenomenon.

Following the system inside coexistence along the line $A-B$, 
the neutron chemical potential increases.
Phase separation implies that, in this region,
the system is composed of two phases located at 
the coexistence boundaries and correponding to the same value of the 
intensive parameters.
The diluted phase goes along coexistence
from A to A' while the dense phase goes on the other side of 
the coexistence border from B' to B. When it reaches B 
the gas is entirely transformed into a liquid, the 
phase transition is over and the $Z/A=0.3$ transformation 
corresponds to a homogenous system again. 

\begin{figure}[tbph]
\begin{center}
\includegraphics*[height=0.8\linewidth]{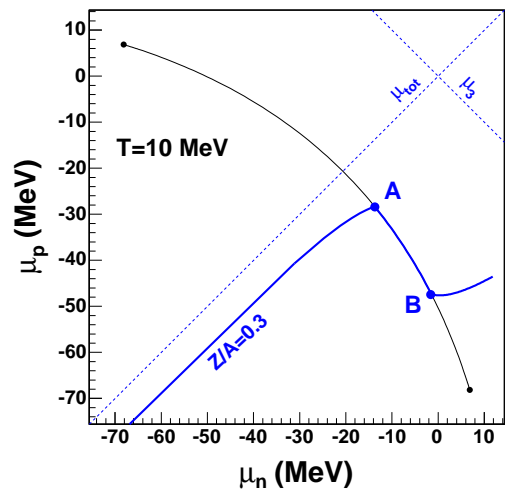}
\end{center}
\caption
	{
Coexistence line, i.e. line of first-order phase transition, 
in the intensive-parameter 
plane $(\mu_n,\mu_p)$. The path at $Z/A=0.3$ is also shown.
	}
\label{MnMp}
\end{figure}

The evolution of $\mu_n$ during the transition can be 
quantitatively discussed on figure 
\ref{MnZsA} which gives a projection of figure \ref{MnMpZsA} on the
$\mu_n$ axis. This figure clearly 
shows that a constant-$Z/A$ transformation does not cross 
the coexistence at a unique $\mu_n$ value but explores
a finite range of chemical potentials. 
The system is thus forced to follow the coexistence line in the intensive 
parameter space, as shown in figure \ref{MnMpP} and in figure \ref{MnMp}. 
The only exceptions are symmetric matter and the 2 maximum asymmetries
of the coexistence region for each temperature.

It is important to notice that 
the behavior shown in figures \ref{MnZsA} and \ref{MnMp} is 
the generic behavior expected when 
a conservation law is imposed on an order parameter \cite{noi-LectureNotes, noi-PhysRevE}. 
Indeed, the usual discontinuity of the 
order parameter characteristic of a first-order transition is 
prevented by the constraint. If the system reaches coexistence, the only 
way to fulfill the conservation law on the order parameter is to 
follow the coexistence line until the conservation law becomes 
compatible with a homogeneous phase. 

\begin{figure}[tbph]
\begin{center}
\includegraphics*[height=1.\linewidth]{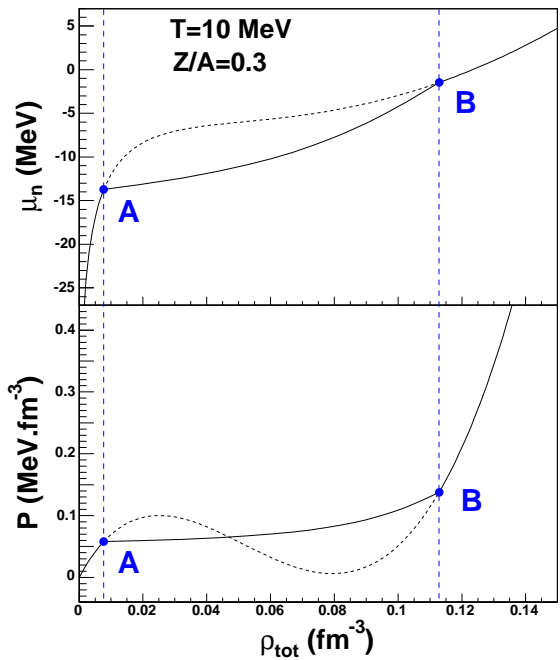}
\end{center}
\caption
	{
Transformation at constant $Z/A=0.3$ in nuclear matter at $T=10MeV$.
Top : $\mu_n(\rho_{tot})$. 
Bottom : $P(\rho_{tot})$.
Dotted lines : solution for an homogeneous system.
Full lines : Gibbs construction. 
Points $A$ and $B$ give the coexistence-region borders. 
	}
\label{rP-rMn}
\end{figure}

This continuous evolution by phase mixing hides the EOS discontinuity associated with
the first-order phase transition.This is illustrated by the evolution of $\mu _{n}$ and $P$ 
as a function of the total density $\rho_{tot} $ at $Z/A=0.3$ in figure \ref{rP-rMn}. 
Full lines are obtained after the thermodynamic
potential is made convex by phase mixing, i.e. with the Gibbs construction. 
These functions present no plateau in the coexistence region, and 
the transformation looks like a continuous transition. 
Yet by definition the system is going through a first-order phase transition
since the first derivatives of the grand potential are discontinuous. 

This clearly stresses the fact that the behavior of specific transformations 
should not be confused with the intrinsic thermodynamic properties.
Indeed a transformation is an explicit or implicit relation between the thermodynamic variables, e.g. a constant proton fraction $Z/A$ implies $\rho_{p}=\rho_{n}Z/(A-Z)$. Then the variation of the intensive variables conjugated to one of the order parameters, e.g $\mu_{n}$ for $\rho_{n}$ is given by :

\begin{equation}
\left(\frac{d\mu_{n}}{d\rho_{n}}\right)_{Z/A=cst}=\frac{\partial \mu_{n}}{\partial \rho_{n}} 
+\frac{\partial \mu_{n}}{\partial \rho_{p}} \frac{Z}{A-Z}
\label{EQ:dmnsdrn}
\end{equation}

since

\begin{equation}
\left(\mu_n(\rho_n)\right)_{Z/A=cst}
=\frac{\partial f}{\partial \rho_{n}} \left(\rho_n,\rho_n\frac{Z}{A-Z}\right)
\end{equation}

Then a first-order phase transition characterized by $\partial \mu_{n}/\partial \rho_{n}=0$ leads to a continuous evolution of $\mu_{n}$ as a function of $\rho_{n}$ according to eq.(\ref{EQ:dmnsdrn}) because of the constraint of constant proton fraction.

The isospin degree of freedom does not change the order of the nuclear 
liquid-gas phase transition as claimed in different 
articles \cite{Muller-Serot, Carmona}. 
It remains first order. Only the constant-proton-fraction transformations 
(or other transformations constraining an order parameter) mimic a continuous 
transition because they do not cross the coexistence line at a single point, 
but are forced to follow it to fulfill the conservation law. 

In general, transformations involving a constraint on an order parameter always appear 
continuous even in the presence of a first order phase transition \cite{noi-LectureNotes, noi-PhysRevE}. 

\subsection{Phase diagram: $T=0$ specificity}

Until now we have presented a study at a fixed finite temperature. 
We will now consider the effect of temperature on the phase diagram, 
from the particular case of zero temperature to the symmetric-matter 
critical temperature $T_c$ above which there is no transition any more.

\begin{figure}[tbph]
\begin{center}
\includegraphics*[width=0.9\linewidth]{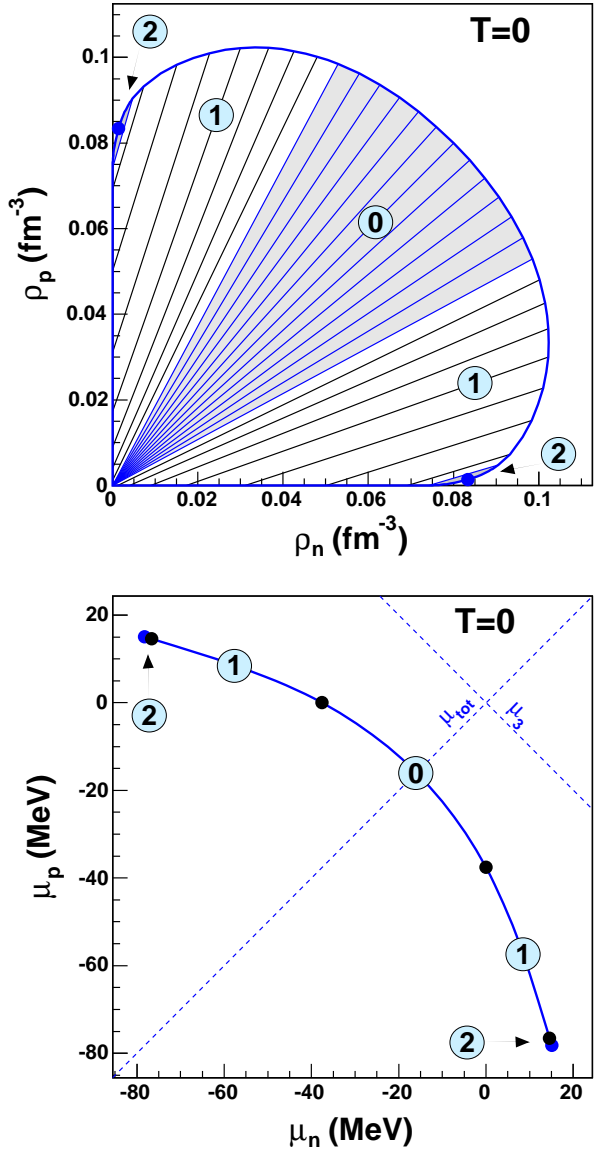}
\end{center}
\caption
	{ 
Coexistence region at T=0. 
It is split into 3 different types of equilibria between liquid and gas (see text).
Top : coexistence border (thick line) in the density plane. 
The straight lines relate two coexisting phases.
Bottom : coexistence line in the plane of chemical potentials.
	}
\label{bulle0-rnrp}
\end{figure}

The specificity of the zero-temperature case 
is the possibility to reach a vanishing
density with a finite chemical potential \cite{Pethick},
while at any finite temperature a given density can be zero 
only if the associated chemical potential goes to $-\infty$ 
(see eq.(\ref{EQ:density})).
This is a trivial consequence of the 
specificity of Fermi-Dirac distribution at $T=0$.
In this case, the associated thermodynamic potential
presents at zero density an edge with a finite slope, 
the associated finite chemical potential. An example is 
given by the (free) energy as a function of total density
for symmetric nuclear matter in figure \ref{FIG:FsV-FsA-T0}.
Then, if the thermodynamic potential for the uniform system 
presents a concave region reaching zero density, the construction of the 
convex envelope does not reduce to the usual tangent construction 
between two points in coexistence. 
Indeed the interpolating plane defined by phase mixing
will not be tangent to the thermodynamic potential
on the zero-density edge.

The coexistence region at $T=0$ is shown in figure \ref{bulle0-rnrp}.
It can be divided into three zones corresponding to
three kinds of equilibria with different conditions on the intensive
parameters. 
In two small regions, labeled $2$ in figure \ref{bulle0-rnrp} and 
associated by isospin symmetry, 
both proton and neutron densities are finite in the gas phase.
The usual bi-tangential construction can then be performed. 
This imposes the standard equality between all intensive parameters 
in the two phases A and B: $\beta(A)=\beta(B)$,
$\mu_{n}(A)=\mu_{n}(B)$, $\mu_{p}(A)=\mu_{p}(B) $ and $P(A)=P(B)$. 
This region is limited on one side by a critical point
beyond which there is no more curvature anomaly, 
and on the other side by the vanishing of one of the two densities 
in the gas phase. 
This second case corresponds to two symmetric regions denoted $1$
in figure \ref{bulle0-rnrp}. There, 
we have equilibria for which the low density phase 
is on an edge $\rho_{q}=0$, the second density being finite. 
For simplicity, let us take the case of globally neutron-rich matter
where $\rho_p=0$ and $\rho_n>0$ at the low-density coexistence edge. 
In this case, the convex envelop is tangent to the thermodynamic 
potential of the uniform system only in the dense phase.
This is a mono-tangential construction. It means that 
phase equilibrium does not require the equality 
of the potential derivative in the $\rho_{p}$ direction, i.e. the 
equality of the proton chemical potentials in the two coexisting phases.
Since the anomaly involves only the derivative in the $\rho_{p}$ direction,
the other equilibrium conditions characterizing the Gibbs construction
are still satisfied in this region, namely the neutron chemical potential,
pressure and temperature have to be the same in the two phases.

\begin{figure}[tbph]
\begin{center}
\includegraphics*[width=0.9\linewidth]{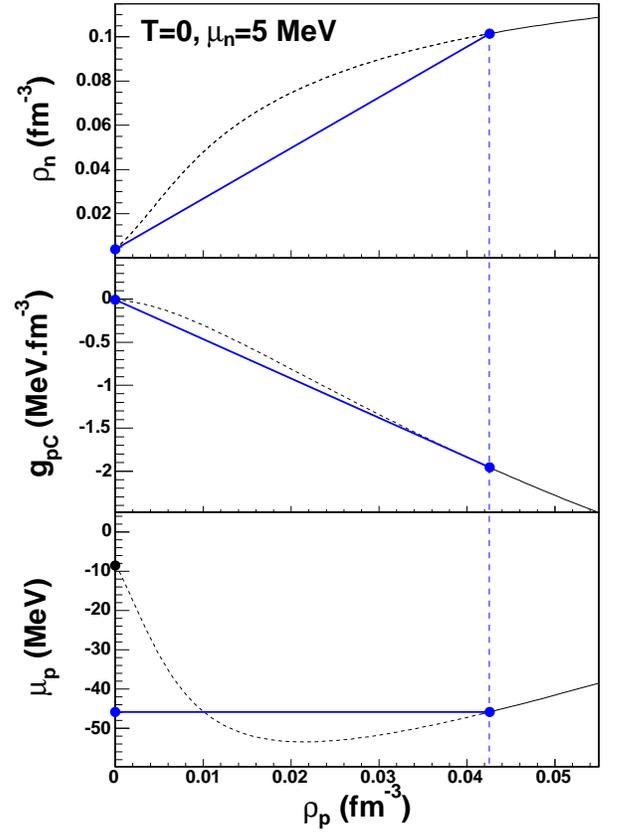}
\end{center}
\caption
	{
Specificity of $T=0$ : 
Mono-tangential construction for a fixed neutron chemical potential $\mu_n=5 MeV$.
Upper part : corresponding path in the density plane. 
It reaches the edge $\rho_p=0$ (with a finite $\rho_n$).
Middle part: thermodynamic potential density $g_{pC}$ (see section III-B). 
The concave region is corrected by phase mixing, 
which corresponds here to a mono-tangential construction (straight line). 
Lower part: $\mu_p$. The mono-tangential construction on $g_{pC}$ does not correspond to the usual Maxwell construction, that can not be performed because of the finite value of $\mu_p$ on the edge. 
	}
\label{isoMn}
\end{figure}

Since the construction involves two points with the same $\mu_{n}$,
we can use the neutron-grand-canonical proton-canonical
potential $G_{pC}$ (see section III-B) in order to reduce 
the problem to a one-dimensional case as illustrated in figure \ref{isoMn}. 
This is a way to solve the problem in practice. 
Looking at a fixed $\mu_{n}$ for which the proton density reaches its 
edge $\rho_{p}=0$ at zero temperature (upper part), we can see that 
a standard equal-area Maxwell construction
is not possible in this case (lower part). However, the concave region 
in the thermodynamic potential (middle part) has to be corrected by phase mixing.
Equilibrium is then given by a mono-tangential construction on $g_{pC}$.

Such equilibrium between a neutron gas and a two-fluid liquid 
leads to a discontinuous change of $\mu_{p}$ in the gas.
Because of the dominance of region $1$ with respect to region $2$ 
in the phase diagram, the simplification is often made in 
the literature \cite{Pethick} that phase coexistence 
in neutron-star matter can be modelized as the equilibrium between 
neutron-rich nuclei and a pure neutron gas. It should however
be noticed that this is possible only if the temperature is exactly
zero, a finite proton fraction being associated to the gas phase 
at any finite temperature.

The last case, denoted $0$ in figure \ref{bulle0-rnrp}, 
corresponds to 
both $\rho_p=0$ and $\rho_n=0$ in the low-density phase.
This case corresponds to a dense phase 
in equilibrium with the vacuum, i.e. at zero pressure. 
This region has a very simple physical interpretation.
If the gas phase is given by the point $(\rho_n,\rho_p)=(0,0)$, 
this means that the coexistence lines of zone $0$ are constant $Z/A$ lines. 
The coexistence border on the liquid side is the
locus of zero pressure in the $Z/A$ interval corresponding 
to region $0$. For each value of $Z/A$, the liquid border 
is then given by the minimum of the energy per particle 
computed for this constant $Z/A$. Zone $0$ corresponds then to the 
chemical potential interval in which a self-bound liquid
can be defined. 

\subsection{Phase diagram: temperature dependence}

\begin{figure}[tbph]
\begin{center}
\includegraphics*[width=0.9\linewidth]{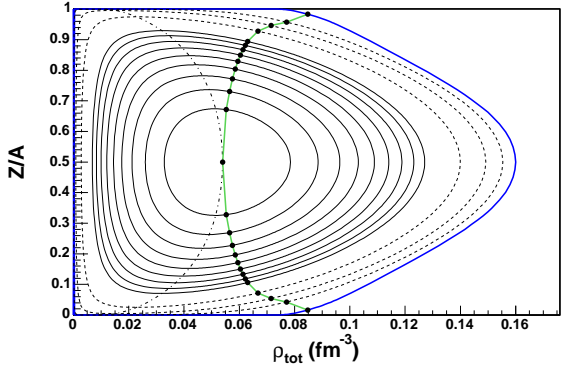} 
\end{center}
\caption
	{ 
Coexistence region in the $(\rho,Z/A)$ plane at different temperatures. 
Thick line : $T=0$.
Dotted lines : $T=4$, $6$, $8 MeV$.
Solid lines : $T=10$, $10.5$, $11$, $11.5$, $12$, $12.5$, $13$, $13.5$, $14 MeV$. 
Black dots : critical points at each temperature, including $T_c=14.54 MeV$. 
Dashed-dotted line : points of maximum asymmetry for coexistence at each temperature.
	}
\label{rZsA}
\end{figure}

In order to determine the phase diagram of nuclear matter, 
we establish the ensemble of points at equilibrium in the $(\rho_n,\rho_p)$ 
plane for different fixed values of temperature. 
The result is represented in figure \ref{rZsA}
as a function of the variables $(\rho,Z/A)$ 
in order to underline the role of isospin. 
Since protons and neutrons play symmetric roles, 
the resulting curves are symmetric with respect to the axis $Z/A=0.5$. 
For this reason, in the following, only the neutron-rich part ($Z/A<0.5$) 
will be discussed.

\begin{figure}[tbph]
\begin{center}
\includegraphics*[width=0.9\linewidth]{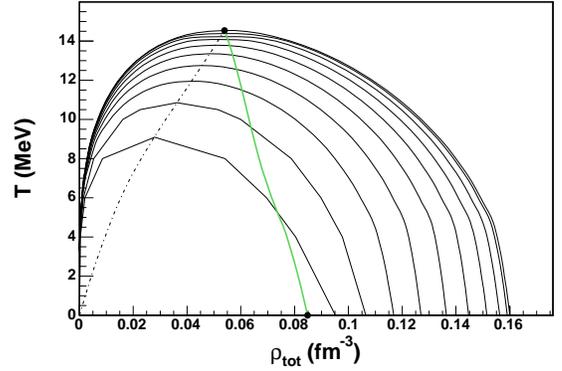} 
\end{center}
\caption
	{
Cuts of the coexistence region as a function of the 
total density for regularly spaced values of Z/A between $0.05$ and $0.5$.
%$Z/A=0.5$, $0.45$, $0.4$, $0.35$, $0.3$, $0.25$, $0.2$, $0.15$, $0.1$ and $0.05$. 
Thick grey line : line of critical points. 
Dashed-dotted line : line of maximum asymmetry.
	 }
 \label{tranches-ZsA}
 \end{figure}

\begin{figure}[tbph]
\begin{center}
\includegraphics*[width=0.9\linewidth]{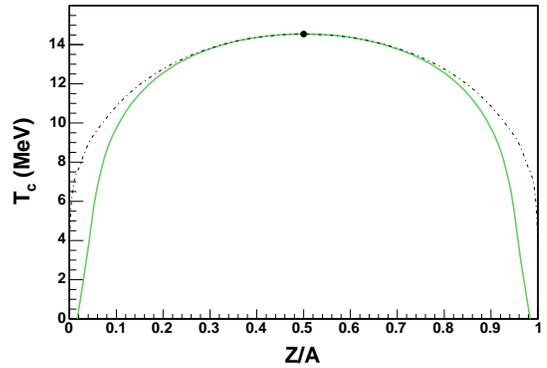}
\end{center}
\caption
	{
Critical temperatures as a function of the system Z/A (thick grey line). The curve of maximal temperatures for coexistence is also shown (dashed-dotted line).
	}
\label{ZsA}
\end{figure}

The widest coexistence region corresponds to zero temperature. 
Equilibrium points that belong to the edge $\rho_p=0$ make a line at $Z/A=0$. 
When temperature is introduced, coexistence region concerns non-zero values
of $\rho$ and $Z/A$, which means that equilibrium is always 
between phases containing both kinds of particles. 

The critical points are also reported for each 
temperature. They are second-order-transition points. 
They form a critical line.
An interesting feature is that the critical density 
increases with the asymmetry 
while the critical temperature decreases, as
can also be seen from figures \ref{tranches-ZsA} and \ref{ZsA}.
This is in agreement with previous studies
using different effective interactions \cite{Muller-Serot, BaoAnLi}. 

It should be stressed that, on each coexistence curve at a fixed temperature,
the critical points do not correspond to the minimum (or maximum) $Z/A$. 
This point $((Z/A)^{min},\rho^{min})$ is situated at a
density lower than the critical point, i.e. on the gas side of coexistence. 
As a result, transformations can be performed at constant $Z/A$ 
such that the system enters at a point of coexistence as a gas and exits 
at another point still as a gas. This is the so-called retrograde
transition \cite{Huang}.
This happens for the small region for which $(Z/A)^{min}<Z/A<Z/A^c$. 
Between those two points, there is appearance and disappearance of a 
liquid phase at $(Z/A)^L>Z/A$ at equilibrium with a gas at $(Z/A)^G<Z/A$. 
For a too neutron-rich system such that $Z/A<(Z/A)^{min}$,
there can be no phase coexistence. 

As temperature grows, the coexistence region is reduced. 
The values of $(Z/A)^c$ and $(Z/A)^{min}$ become closer 
to the symmetry $Z/A=0.5$. 
Densities $\rho^c$ and $\rho^{min}$ follow opposite evolutions: 
$\rho^{min}$ grows with temperature, while $\rho^c$ diminishes. 
The coexistence region disappears at the critical temperature $T^c$, 
for which only symmetric matter presents a second-order transition point. 
Critical and minimum Z/A lines join at this ultimate critical point.

It is also interesting to look at the 
coexistence zone as a function of $Z/A$
(see figures \ref{tranches-ZsA}, \ref{ZsA}).
We can see that the dependence of the critical temperature (as well as the maximal temperature)
on $Z/A$ is weak. Only for very high values of asymmetry the difference
between the two temperatures for a fixed
$Z/A$ can be several MeV. This is another way to visualize 
the retrograde transition, meaning that we can have saturated
vapor at supercritical temperatures.

\begin{figure}[htbp]
\includegraphics*[width=0.9\linewidth]{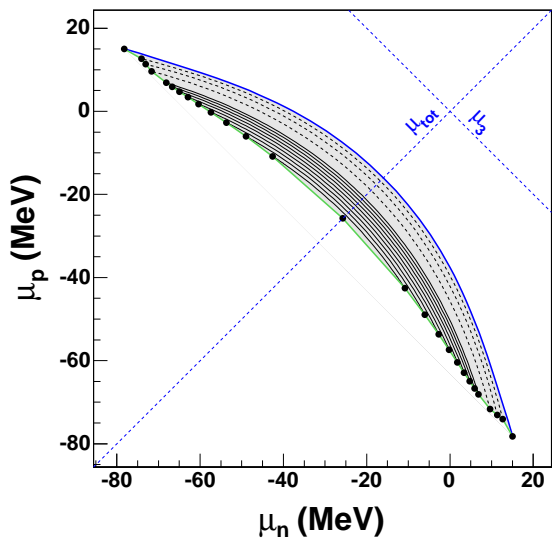}
\caption	
	{
Phase diagram in $(\mu_n,\mu_p)$ for different values of $T$. 
Thick line : $T=0$. 
Dotted lines : $T=4$, $6$, $8 MeV$. 
Solid lines : $T=10$, $10.5$, $11$, $11.5$, $12$, $12.5$, $13$, $13.5$, $14 MeV$. 
Black dots : critical points at each temperature, including $T_c=14.54 MeV$.
}
\label{17}
\end{figure}

Finally, it is interesting to look at the coexistence manifold in the 
$(\beta,\mu_n,\mu_p)$ space, as shown in figure \ref{17}. 
One can see that the coexistence is
almost perpendicular to the $\mu_3=0$ axis, stressing the fact that the
nuclear liquid-gas phase transition is dominated by 
its isoscalar component \cite{Margueron, Baran-PhysRep}. 

\subsection{Critical behavior}

Let us now study in more details the critical behavior of the system. 
Conversely to the usual single-fluid liquid-gas phase transition which presents 
a unique critical point, a two-fluid system is critical along a line in the 
($\beta$, $\mu_n$, $\mu_p$) intensive-parameter space. The critical 
points are obtained for each temperature (below the symmetric-matter 
critical temperature) by determining the $\mu_n$ value for
which the two phases with the same tangent plane merge together.
To evaluate this point, we have used both the Maxwell construction 
in the neutron-grand-canonical proton-canonical ensemble (see section III-B) 
and the existence of a crossing point in the $P$ versus $\mu_p$ 
curve at constant $(T,\mu_n)$. 
The critical points also correspond to the disappearance of a concave 
region in the uniform-system free energy. This disappearance of the 
spinodal region provides an independent way to evaluate the location of the 
critical line:
we can represent as a function of $\mu_n$ the lower value 
of the free-energy curvature in the $(\rho_n,\rho_p)$ plane. 
The resulting curve crosses the horizontal axis at the 
critical value of $\mu_n$. 
Below the symmetric-matter critical point $T_c$, these methods lead 
to the definition of two symmetric critical lines,
one for neutron-rich systems 
($\mu_n=\mu_n^>(T)$, $\mu_p=\mu_p^<(T)$ )
and the other for the opposite isospin 
($\mu_p=\mu_p^>(T)$, $\mu_n=\mu_n^<(T)$ ).

\begin{figure}[tbph]
\begin{center}
\includegraphics*[width=0.9\linewidth]{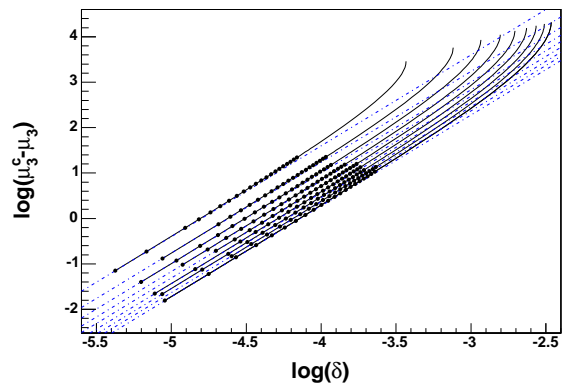} 
\end{center}
\caption
	{	
Solid lines with points : evolution of $\mu_3-{\mu_3}_c$ as a function of 
$\delta$ (see text) near the critical point for different temperatures : 
$T=10$, $10.5$, $11$, $11.5$, $12$, $12.5$, $13$, $13.5$, $14 MeV$.
Straight dashed-dotted lines : power laws with critical exponent $\beta_{\beta}=2$.
 	}
 \label{lnd-lnM3}
 \end{figure}

\begin{figure}[tbph]
\begin{center}
\includegraphics*[width=0.9\linewidth]{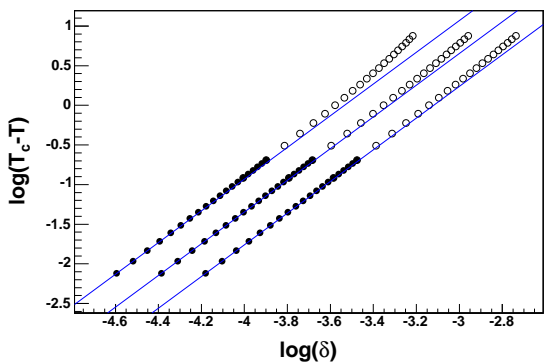}
\end{center}
\caption
	{
Points : evolution of $T-T_c$ as a function of $\delta$ (see text) 
near the critical point for different $\mu_n$: 
$\mu_n=\mu_q^<(T_c)$, $\mu_q^<(T=12 MeV)$, $\mu_q^<(T=10 MeV)$.
Straight lines: power laws with critical exponent $\beta_{\mu_n}=2$.
 	}
 \label{lnd-lnM3lnT}
 \end{figure}

At the approach of the critical point, the distance in the space of
observables between the two phases ($A$ and $B$) in equilibrium, 
i.e. the order parameter, goes to zero 
as a power of the distance to the critical point in the 
intensive-variable space. 
The resulting power law is characterized by the critical exponent
$\beta$.
In order to study this behavior for nuclear matter in the 
($\rho_n,\rho_p$) plane, we consider a distance 
$\delta=\sqrt{(\rho_n^B-\rho_n^A)^2+(\rho_p^B-\rho_p^A)^2}$.
In a two-fluid system the behavior of this distance can be studied in two 
ways. 

First, for a fixed temperature $T=\beta^{-1}$,
 $\delta$ is expected to depend on the chemical potentials as :

\begin{equation}
\delta_{\beta}(\mu_q)\propto(\mu_q^c-\mu^q)^{1/\beta_{\beta}}
\end{equation}

where $\beta_{\beta}$ is
the critical exponent at a fixed temperature and $\mu_q^c$
the critical value of $\mu_q$ for the considered temperature. 
Since $\mu_n$ and $\mu_p$ are constrained to be on the 
coexistence line, the above relation is independent of the chemical 
potential selected to perform the study.
For symmetry reasons we have introduced the isoscalar 
$\mu=\mu_n+\mu_p$ and the isovector $\mu_3=\mu_n-\mu_p$ 
chemical potentials. 
Since the dependence on $\mu$ is negligeable (see figure \ref{17}), we have focused our study on the evolution of $\delta$ with $\mu_3$.
The results are presented in figure \ref{lnd-lnM3}.
We can observe that our results perfectly follow 
the expected critical behavior 
with the mean-field value $\beta_{\beta}=2$. 

For a fixed value of $\mu_q$, the distance $\delta_{\mu_q}$ 
goes to zero as the temperature tends to the critical one 
$\beta^c $. The power law should be :

\begin{equation} 
\delta_{\mu_q}(\beta)\propto(\beta^c-\beta^q)^{1/\beta_{\mu_q}}
\end{equation}

where $\beta_{\mu_q}$ is the critical exponent at a fixed $\mu_q$. 
Again, since the temperature $\beta$ and the chemical potential $\mu_q$
in coexistence are related by a Clapeyron-like relation, the above study 
leads to the same critical behavior if studied as a function of 
$\mu_{q'}$ with $q' \neq q$, instead of $\beta$.
In figure \ref{lnd-lnM3lnT} we present the evolution with temperature 
for different chemical potentials. 
This graph shows that our results fulfill the
mean-field scaling $\beta_{\mu_q}=2$. 

\section{Conclusion}

In this paper, based on a mean-field analysis of nuclear matter 
with a realistic Skyrme SLy230a effective interaction, we have established 
that nuclear matter presents a first-order 
phase transition even when the isospin degree of freedom 
is explicitly accounted for. 
This results from the existence of a spinodal region, 
which is a region where the
free energy of a homogeneous system is concave. 
In the case of infinite systems,
such curvature anomaly is corrected by constructing the convex envelope.
This is a tangent construction that links pairs of points 
in the space of observables with the same values for the 
intensive parameters. 
It corresponds to points of discontinuous first 
derivatives for the grand-canonical potential of the system in the space of Lagrange intensive parameters, along a coexistence manifold limited by a critical line.
Except on this limit which 
corresponds to a second-order phase transition, 
the slope discontinuity demonstrates that 
the system is undergoing a first-order phase transition. 
For fixed values of temperature (below the symmetric-matter critical one), 
coexistence lines are obtained in the chemical-potential plane. 
They are limited by critical points that correspond to proton fractions depending on
the temperature. As lower temperatures are considered, more asymmetric nuclear
matter can be involved in a first-order phase transition.

As for the study of critical behaviors, we have found that all the 
numerical data can be fitted by power laws with critical exponents equal to $2$, 
which is consistent with generic mean field predictions \cite{Stanley}.
Calculations beyond mean field are needed 
to obtain a result which could be characteristic of a given universality
class. 
Looking at the isospin content of the phases, we show 
that the proton fraction is discontinuous at the phase 
transition except at the critical points and for symmetric matter. 
In asymmetric nuclear matter, the proton fraction can be used as 
an order parameter. When the transition occurs, the liquid gets closer 
to symmetry while the gas is enriched in the more abundant species. 
This is the well-known isospin fractionation. 
This has a strong influence on the constant-proton-fraction 
transformations, since 
in order to fulfill the imposed conservation on an order 
parameter, the transformation is forced to follow the coexistence 
line instead of crossing it. 
This hides the slope discontinuity characteristic of a first-order
phase transition, the transformation 
appearing as continuous. 

%This study is to be followed by several developments. 
%Within the same framework of nuclear matter in mean field, 
%more characteristics of the transition can be worked out by 
%considering isospin fractionation between liquid and gas, 
%in and and out of equilibrium. 
%About critical exponents, it is necessary to go beyond mean field 
%to obtain a result which could be characteristic of a given universality class.
%This can be don by using a lattice-gas model. 
%Finally the discussion about the order of the transitions can be 
%transposed in the case of finite systems, studying curvature anomalies 
%in the entropy. Then, both 
%Coulomb and the finite size effects can be included. 
%This study can also be related
%to supernovae and neutron stars physics 
%including electrons and neutrinos. 

\end{document}